\documentclass[conference]{IEEEtran}
\IEEEoverridecommandlockouts
\usepackage{cite}
\usepackage{amsmath,amssymb,amsfonts}
\usepackage{algorithmic}
\usepackage{graphicx}
\usepackage{textcomp}
\usepackage{nomencl}
\usepackage{xcolor}
\usepackage{caption}
\usepackage{bm}
\usepackage{tabularx}
\usepackage{cancel}
\usepackage{subcaption}
\usepackage{threeparttable}
\usepackage{siunitx}
\usepackage{etoolbox}
\usepackage{makecell}
\usepackage{float}
\renewcommand\nomgroup[1]{%
  \item[\bfseries
  \ifstrequal{#1}{S}{List of Subscripts}{%
  \ifstrequal{#1}{P}{List of Symbols}{}%
  \ifstrequal{#1}{A}{List of Abbreviations}{}%
  }]%
}

\newcommand{\bld}[1]{\mbox{\boldmath $#1$}}

\makenomenclature

\nomenclature[A]{\textbf{EMLA}}{Electromechanical linear actuator}
\nomenclature[A]{\textbf{EHLA}}{Electrohydraulic linear actuator}
\nomenclature[A]{\textbf{HDMM}}{Heavy-duty mobile manipulator}
\nomenclature[A]{\textbf{PMSM}}{Permanent magnet synchronous motor}
\nomenclature[A]{\textbf{RNEA}}{Recursive Newton-Euler algorithm}
\nomenclature[A]{\textbf{TCP}}{Tool center point}
\nomenclature[A]{\textbf{OHM}}{Off-highway machines}
\nomenclature[A]{\textbf{DoF}}{Degree-of-freedom}
\nomenclature[A]{\textbf{ICE}}{Internal combustion engine}
\nomenclature[A]{\textbf{EV}}{Electric vehicle}
\nomenclature[A]{\textbf{PD}}{Partial derivative}
\nomenclature[A]{\textbf{EM}}{Electric motor}
\nomenclature[S]{$UVW$}{Indicating the phase of electric motor}
\nomenclature[S]{$m$}{Indicating the parameters related to electric motor}
\nomenclature[S]{$d$,$q$}{Related to the d- and q-axis}
\nomenclature[S]{$eq$}{Equivalent value}
\nomenclature[S]{$p$}{Indicating the parameters related to planetary gear}
\nomenclature[S]{$g$}{Indicating the parameters related to gearbox}
\nomenclature[S]{$s$}{Indicating the parameters related to screw mechanism}
\nomenclature[S]{$l$}{Indicating the parameters related to load side}
\nomenclature[P]{$\bld{\Psi}$}{Sensitivity metrics}
\nomenclature[P]{$\eta$}{Efficiency of the mechanism}
\nomenclature[P]{$n_p$}{Number of pole pairs}
\nomenclature[P]{$j$}{Inertia (\SI{}{\kilogram\cdot\meter\squared})}
\nomenclature[P]{$m$}{Mass (\SI{}{\kilogram})}
\nomenclature[P]{$k$}{Stiffness (\SI{}{\newton\per\meter}) or (\SI{}{\newton\cdot\meter\per\radian})}
\nomenclature[P]{$n_{ph}$}{Number of electric motor phases}
\nomenclature[P]{$R$}{Stator resistance of electric motor (\SI{}{\ohm})}
\nomenclature[P]{${S_n}$}{Switch commands (n=$1$, $2$, ..., 2$n_{ph}$)}
\nomenclature[P]{$V_{dc}$}{DC voltage of the inverter (\SI{}{\volt})}
\nomenclature[P]{$V$}{Voltage of the electric motor (\SI{}{\volt})}
\nomenclature[P]{$v$}{Linear velocity (\SI{}{\meter\per\second})}
\nomenclature[P]{$\omega$}{Angular velocity (\SI{}{\radian\per\second})}
\nomenclature[P]{$\rho$}{Screw lead (\SI{}{\meter})}
\nomenclature[P]{$F$}{Force (\SI{}{\newton})}
\nomenclature[P]{$\tau$}{Torque (\SI{}{\newton\cdot\meter})}
\nomenclature[P]{$N$}{Ratio}
\nomenclature[P]{$L$}{Stator inductance of electric motor (\SI{}{\henry})}
\nomenclature[P]{$e$}{Back EMF of electric motor (\SI{}{\volt})}
\nomenclature[P]{$p$}{Number of pole pairs}
\nomenclature[P]{$i$}{Stator current of electric motor (\SI{}{\ampere})}
\nomenclature[P]{$\psi_\textit{PM}$}{Magnetic flux of permanent magnet (\SI{}{\weber})}
\nomenclature[P]{$b$}{Viscous friction (\SI{}{\newton\cdot\second\per\meter}) or (\SI{}{\newton\cdot\meter\cdot\second\per\radian})}
\usepackage{svg}
\def\BibTeX{{\rm B\kern-.05em{\sc i\kern-.025em b}\kern-.08em
    T\kern-.1667em\lower.7ex\hbox{E}\kern-.125emX}}

\begin{document}

\title{System-Level Performance Metrics Sensitivity of an Electrified Heavy-Duty Mobile Manipulator\\
\thanks{This work was supported by Business Finland partnership project "Future all-electric rough terrain autonomous mobile manipulators'' (Grant 2334/31/2022).}}


\author{
    Mohammad Bahari$^{*}$, Alvaro Paz, and Jouni Mattila\\
    Faculty of Engineering and Natural Sciences, Tampere University, 33720, Finland\\
    *Email: mohammad.bahari@tuni.fi
}

\newpage
\thispagestyle{empty} 
    © 2024 IEEE. Personal use of this material is permitted. Permission from IEEE must be obtained for all other uses, including reprinting/republishing this material for advertising or promotional purposes, collecting new collected works for 
    resale or redistribution to servers or lists, or reuse of any copyrighted component of this work in other works. This 
    work has been submitted to the IEEE for possible publication. Copyright may be transferred without notice, after which this
    version may no longer be accessible.
\newpage 

\maketitle

\begin{abstract}
The shift to electric and hybrid powertrains in vehicular systems has propelled advancements in mobile robotics and autonomous vehicles. This paper examines the sensitivity of key performance metrics in a electrified heavy-duty mobile manipulator (HDMM) driven by electromechanical linear actuators (EMLAs) powered by permanent magnet synchronous motors (PMSMs). The study evaluates power delivery, force dynamics, energy consumption, and overall efficiency of the actuation mechanisms. By computing partial derivatives (PD) with respect to the payload mass at the tool center point (TCP), it provides insights into these factors under various loading conditions. This research aids in the appropriate choice or design of EMLAs for HDMM electrification, addressing actuation mechanism selection challenge in vehicular system with mounted manipulator and determines the necessary battery capacity requirements.
\end{abstract}
\printnomenclature
\section{Introduction}
\subsection{Background and Motivations}
\begin{figure}[ptb]
  \centering
   \centerline{{\includegraphics[width=7.5cm]{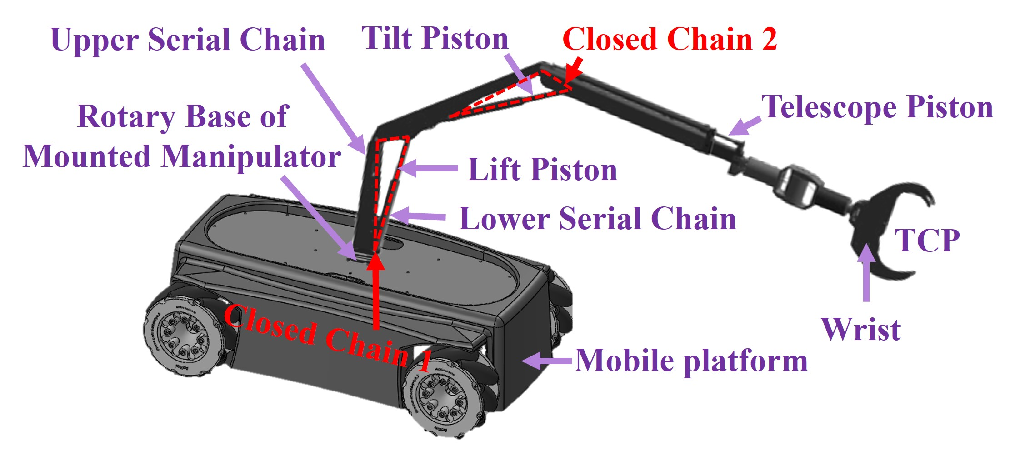}}}
    \caption{Illustrative configuration of an electrified HDMM} 
    \label{fig:Mobile_platform}
\end{figure}

The growing urgency to address climate change and environmental degradation, driven by greenhouse gases like $\text{CO}_2$, has led to a global shift towards sustainability. International agreements such as the Paris Agreement 2015 and the EU's planned ban on ICE vehicle sales by 2035 underscore the need for emission reduction and greener alternatives \cite{agreement2015paris}. Transitioning to electric power, especially from renewable sources, offers an effective path to not only mitigating environmental impacts but also enhances operational efficiency.
In the off-highway machines (OHMs) and heavy machinery sector, traditionally reliant on ICEs, electrifying powertrains represents a major technological advancement \cite{beltrami2021electrification}. With a growing focus on sustainability, the shift towards electric vehicles (EVs), including OHMs, is accelerating due to several key benefits:
\begin{itemize}
\item \textit{Substantial Emission Reductions}: Replacing ICEs with electric alternatives lowers emissions, aiding in global  $\text{CO}_2$ reduction and compliance with environmental regulations, especially in heavy-duty applications.
\item \textit{Superior Energy Efficiency}: Electric powertrains are more energy-efficient than traditional ICEs, leading to reduced energy use and costs, crucial for HDMMs in energy-intensive fields like construction and mining.
\item \textit{Seamless Integration with Automation Technologies}: Electric powertrains’ precision and control enable integration with advanced automation systems, fostering the development of intelligent HDMMs that enhance safety and productivity by automating complex tasks.
\item \textit{Reduced Maintenance/Operational Downtime}: With fewer moving parts than ICEs, electric powertrains require less maintenance and have lower operational downtime, beneficial for HDMMs working remotely.
\end{itemize}
Electrification is becoming crucial for the heavy machinery industry, especially for mobile working machines, which are key in sectors like manufacturing, logistics, agriculture, and search and rescue operations \cite{un2022off}. Integrating electric actuators aligns with sustainability goals and advances intelligent industrial systems. Fig. \ref{fig:Mobile_platform} shows a conceptual setup of an electrified HDMM with a mobile platform and mounted robotic manipulator, demonstrating the shift towards sustainable and autonomous heavy-duty operations.
%
Two main options for electric actuators are electromechanical linear actuators (EMLAs) and electrohydraulic linear actuators (EHLAs). EHLAs offer high force and power-to-weight ratios but face challenges like control limitations, inefficiency from fluid incompressibility, and leakage \cite{jose2021early}. In contrast, EMLAs provide greater precision and efficiency with programmable features that enhance control and reduce maintenance. They avoid energy conversion losses inherent in hydraulic systems, making them a greener choice \cite{10199841}. For a detailed comparison, see Table \ref{tab:EHLA_EMLA} \cite{ratzinger2021electrified}.
%
\begin{table}[tbp]
    \centering
    \caption{Comparing different factors of EHLA and EMLA actuation technology}
    \begin{tabular}{l|c|c}
        \Xhline{2\arrayrulewidth}
        \textbf{Characteristic} & \textbf{EHLA} & \textbf{EMLA} \\
        \Xhline{2\arrayrulewidth}
        Energy efficiency & Moderate & High \\
        \hline
        Environmental impact & High & Low \\
        \hline
        Load capacity & High & Moderate \\
        \hline
        Temperature resilience & Low & High \\
        \hline
        Controllability & Moderate & High \\
        \hline
        Maintenance requirements & High & Low \\
        \hline
        Initial cost & Moderate & High \\
        \hline
        Overload tolerance & High & Moderate \\
        \hline
        Operation sound noise & Moderate & Quite \\
        \Xhline{2\arrayrulewidth}
    \end{tabular}
    \label{tab:EHLA_EMLA}
\end{table}
With the benefits of electrification and advancements in power electronics, driver technology, and electric motors, EMLAs powered by permanent magnet synchronous motors (PMSMs) are expected to significantly impact green vehicular systems, especially for HDMMs since the adoption of PMSM lead to high torque density, high efficiency, better controllability, and low cogging torque \cite{tootoonchian2016cogging,heydari2024robust}, making them ideal for HDMM applications. Ongoing research aims to optimize these systems to meet the rigorous demands of modern industrial practices.
%
%
\subsection{Contributions and Structure of the Paper}
This paper presents a comprehensive system-level sensitivity analysis of a heavy-duty mobile manipulator (HDMM) powered by electromechanical linear actuators (EMLAs) driven by permanent magnet synchronous motors (PMSMs). We develop a detailed model of the EMLA mechanism and generate a trajectory task for the robotic manipulator to ensure accurate tracking of the tool center point (TCP). The analysis evaluates various critical factors, including power delivery, force dynamics in the joint pistons, energy expenditure, and overall efficiency of the EMLAs. Additionally, we compute partial derivatives (PD) with respect to the payload mass at the TCP to examine how these factors vary under different loading conditions throughout the duty cycle. This study offers valuable insights into the performance and behavior of the EMLAs under diverse operational scenarios, aiding in the intelligent design and selection of EMLAs for each joint of the manipulator. Furthermore, it provides guidance for determining the necessary battery capacity to support the electrification of HDMMs. The organization of the paper is:
\begin{itemize}
    \item The research begins with Section \ref{sec:equivalent}, where we develop a detailed model of a gear-equipped PMSM-powered EMLA to analyze energy conversion processes. This model includes equivalent circuits for the inverter, EM, gearbox, screw mechanism, and load. Using the force-to-voltage and speed-to-current analogies, we provide a framework for understanding the governing dynamic of EMLA and analyzing its performance.
    \item In Section \ref{sec:motion}, the method for generating motion for the studied EMLA-driven HDMM is described in detail, as presented in \textbf{Algorithm 1}. This second-order inverse differential kinematics algorithm outlines the process for generating motion. The algorithm updates the payload value in the $robot(\cdot)$ data structure, performs time loop calculations for differential kinematics, computes the robot's inverse dynamics, and evaluates the sensitivity metrics ($\bld{\Psi}$) based on linear velocities ($v_{x_i}$) and forces of the actuators ($f_{x_i}$) at the EMLA's load side.
    \item Section \ref{sec:simulation} presents the simulation and results of the sensitivity analysis on the 3-DoF EMLA-actuated HDMM, providing comprehensive insights into the behavior of the system with respect to the payload and pave the way for appropriate actuator sizing and determining required battery capacity for the HDMM.
\end{itemize}
\section{Modeling and Analysis of EMLA Mechanism}
\label{sec:equivalent}
\begin{figure}[b]
  \centering
   \centerline{{\includegraphics[width=6.5cm]{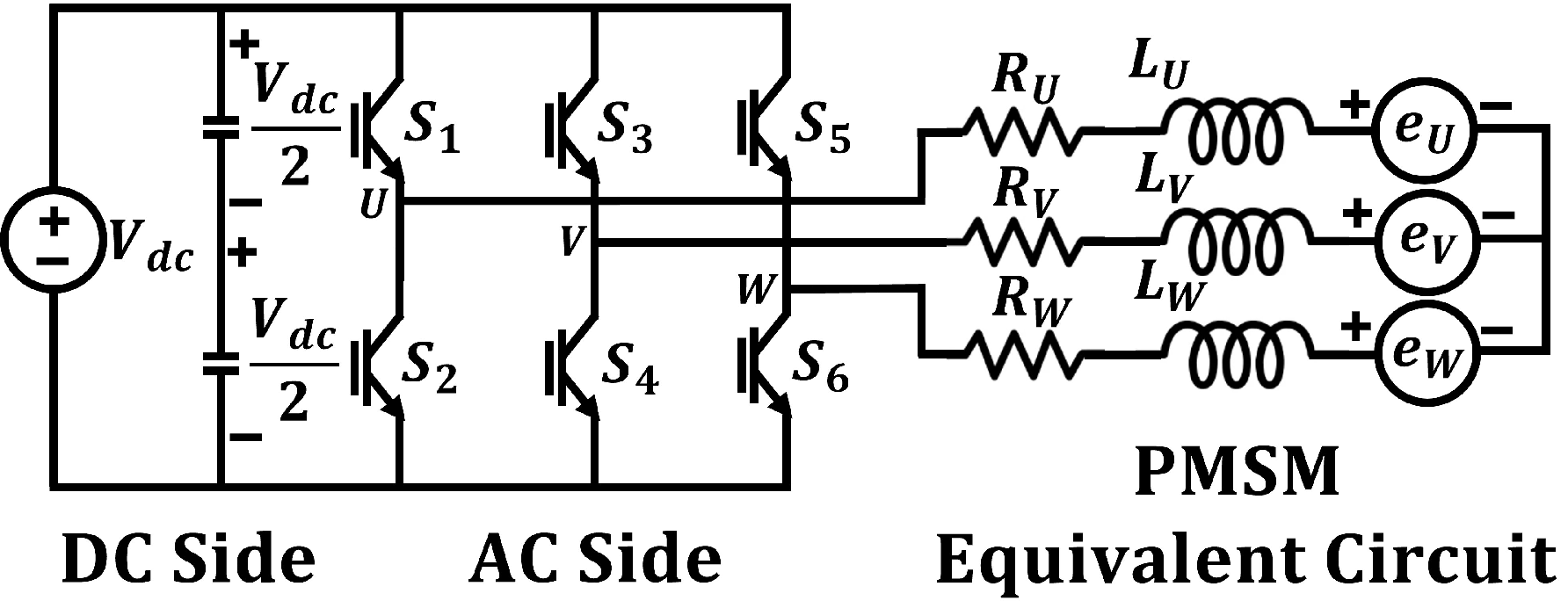}}}
     \caption{Equivalent circuit of 3-phase PMSM} 
    \label{fig:Inverter_pmsm}
\end{figure}

\begin{figure}[t]
  \centering
   \centerline{{\includegraphics[width=7.5cm]{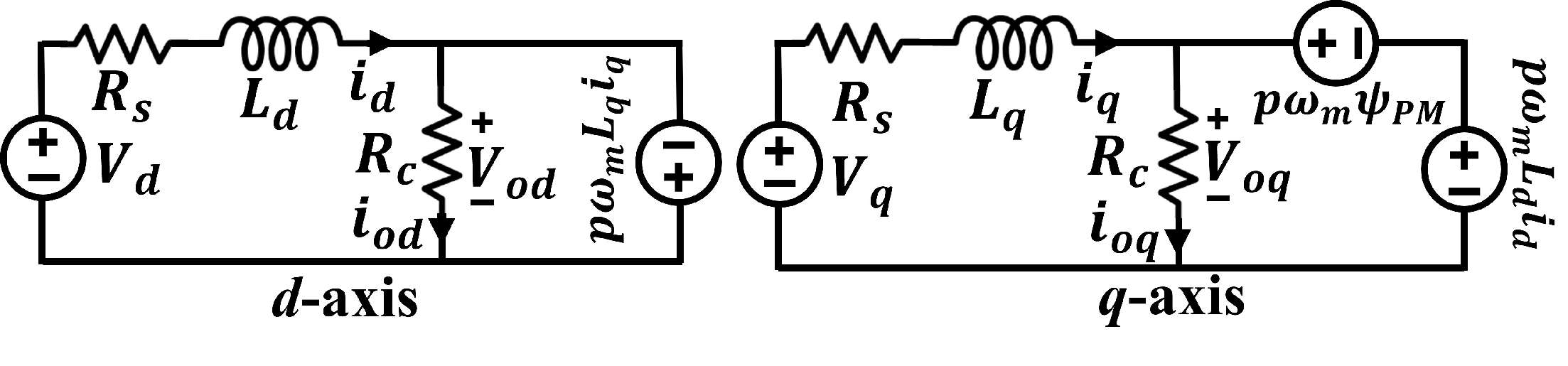}}}
     \caption{Equivalent circuit of PMSM in dq-axis} 
    \label{fig:dq_equivalent}
\end{figure}

\begin{figure}[H]
  \centering
   \centerline{{\includegraphics[width=5cm]{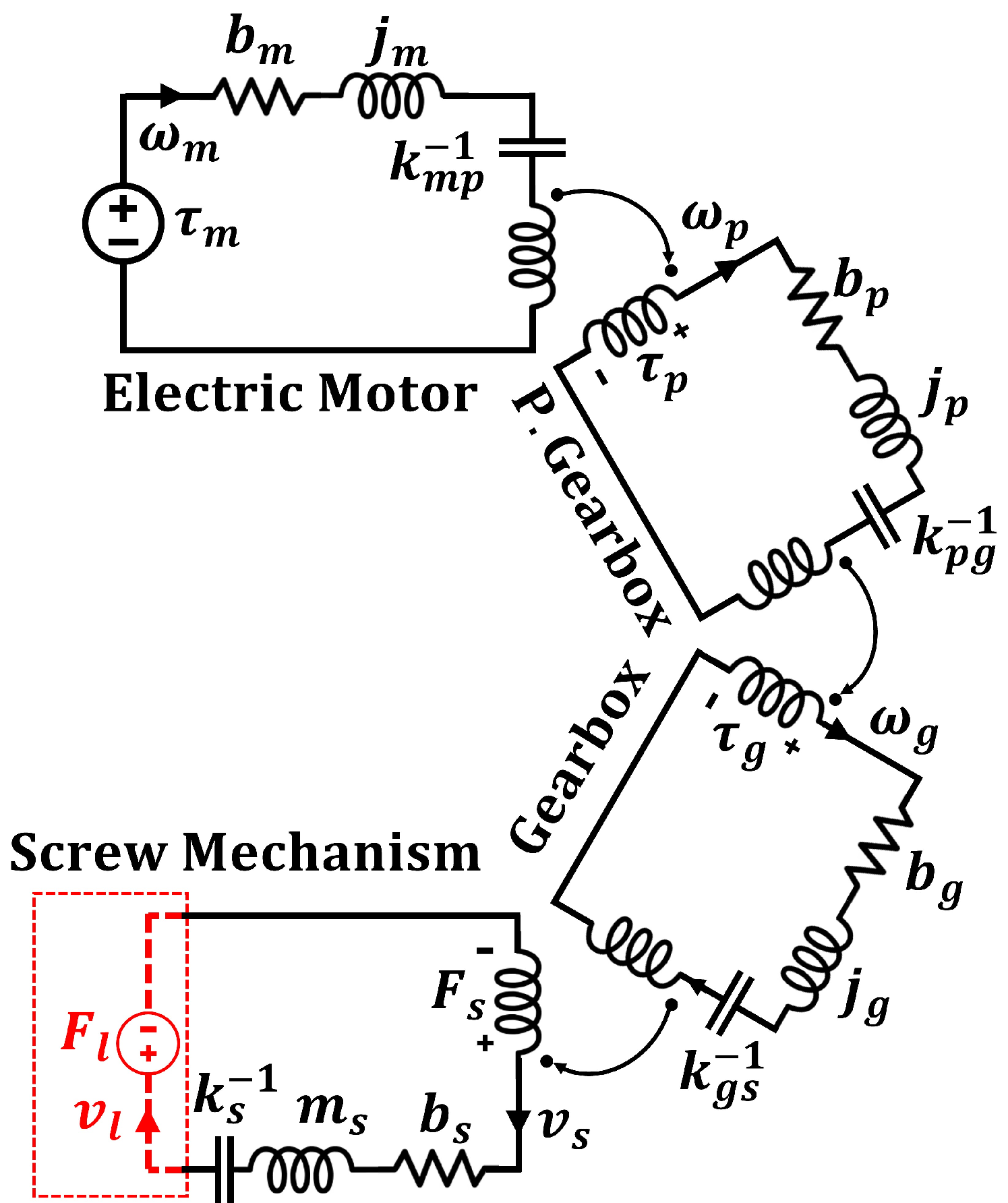}}}
     \caption{Equivalent circuit of mechanical components of EMLA} 
    \label{fig:EMLA_equivalent_circuit}
\end{figure}

\begin{figure}[H]
  \centering
   \centerline{{\includegraphics[width=8.5cm]{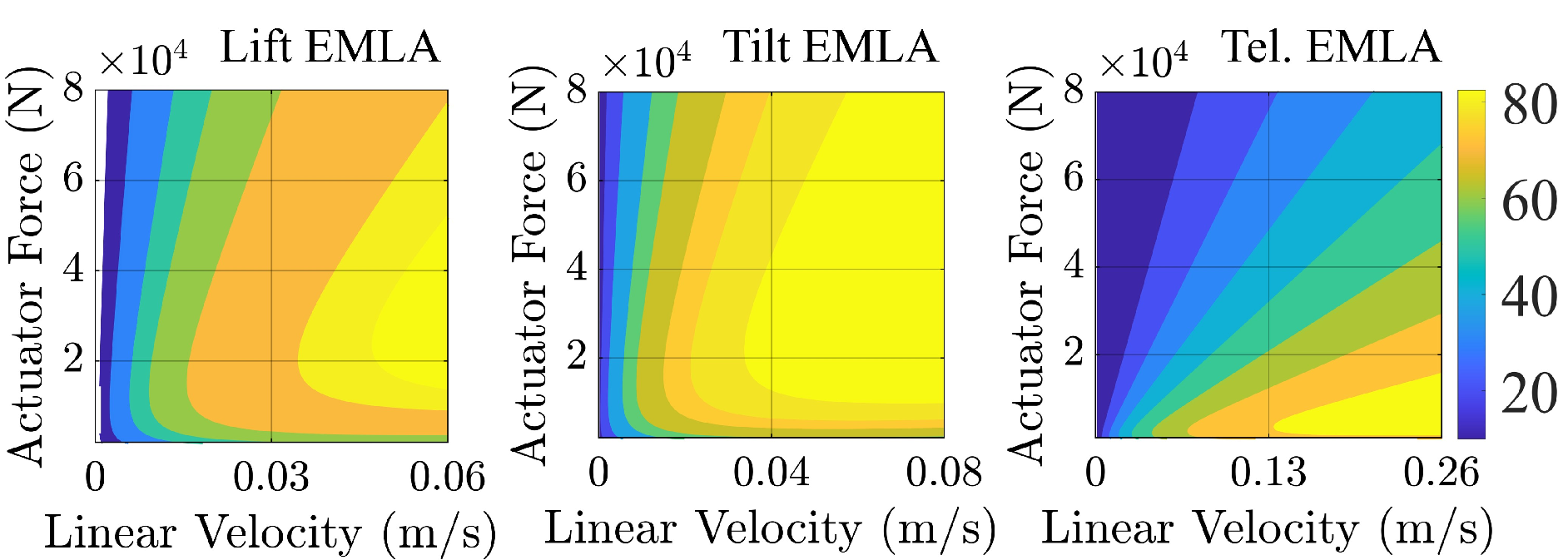}}}
     \caption{Efficiency maps of EMLAs implemented in the lift, tilt, and telescope joints of the studied HDMM} 
    \label{fig:EMLAs_Efficiency}
\end{figure}

In this paper, we develop the model of a gear-equipped PMSM-powered EMLA mechanism (see Fig. \ref{fig:EMLA_schematic}) to analyze governing dynamics. The model includes equivalent circuits of the inverter, motor, gearbox, screw mechanism, and load. We use force-to-voltage and speed-to-current analogies to explore EMLA dynamics. By applying Park’s transformation to the voltage equations and converting them to the $dq$ reference frame, the equivalent circuits of the inverter-driven PMSM are obtained as shown in Fig. \ref{fig:Inverter_pmsm} and Fig. \ref{fig:dq_equivalent}. The voltages ($V_d$ and $V_q$) and the electromagnetic torque are expressed in \eqref{equation:dq0voltage} \cite{krishnan2017permanent}:
\begin{figure}[b]
  \centering
   \centerline{{\includegraphics[width=5.25cm]{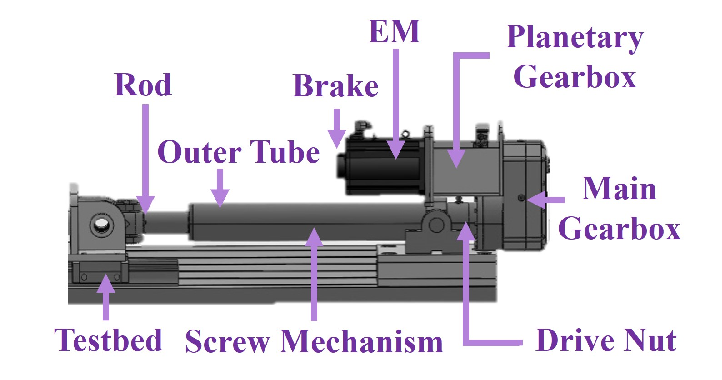}}}
     \caption{EMLA structure and components} 
    \label{fig:EMLA_schematic}
\end{figure}
\begin{equation}
\left\{
\begin{aligned}
V_d &= R_s i_d + L_d \frac{d i_d}{dt} - p \omega_m L_q i_q \\
V_q &= R_s i_q + L_q \frac{d i_q}{dt} + p \omega_m i_d L_d + p \omega_m \Psi_{P\!M}\\
\tau_m &= \frac{3}{2} n_p \left[i_q\left(i_d L_d+\Psi_{P\!M}\right)-i_d i_q L_q\right]
\end{aligned}
\right.
\label{equation:dq0voltage}
\end{equation}
%
By obtaining the equivalent circuit of the EMLA, as depicted in Fig. \ref{fig:EMLA_equivalent_circuit}, the dynamics of the mechanism can be captured. The general expression for the motor torque and EMLA dynamics are provided in \eqref{equation:angularvelocity} to integrate the mechanical part of the PMSM into the rest of EMLA components:
\begin{equation}
\left\{
\begin{alignedat}{6}
&\tau_m      &&= a_{eq} \ddot{x}_L + b_{eq} \dot{x}_L + c_{eq} {x}_L + F_{l}\\
&a_{eq}      &&= \alpha^2 \left(\frac{m_s}{\alpha^2}+ j_{g} + N_g^2 j_p + (N_g N_p)^2 j_{m}\right)\\
&b_{eq}      &&= b_s + \left(\alpha N_g N_p \right)^2 b_{m}\\
&c_{eq}      &&= \alpha^{2} \bigg(\frac{1}{(N_p N_g)^2 k_{mp}} + \frac{1}{(N_g)^2 k_{pg}} + \frac{1}{k_{gs}} + \frac{1}{k_{l}}\bigg)^{-1}\\
&\alpha      &&= \frac{2\pi}{\rho}
\end{alignedat}
\right.
\label{equation:angularvelocity}
\end{equation}
%
 Also, efficiency maps of EMLAs in the manipulator joints can be developed (see Fig. \ref{fig:EMLAs_Efficiency}) by calculating power losses \cite{9269466}.
\section{Sensitivity of Performance Metrics in HDMM}
\subsection{Manipulator Motion Generation}
\label{sec:motion}
Our study focuses on heavy-duty parallel serial HDMM powered by EMLAs, as depicted in Fig. \ref{fig:Mobile_platform}, capable of high-power tasks where payload analysis is crucial for actuator selection. The robot's configurational state is defined in its actuators' space as $\bld{q} \in \mathcal{R}^{n}$ where $n$ is the number of DoF and number of mechanisms in the manipulator.
Motion generation is performed via second-order inverse differential kinematics, as coded in \textbf{Algorithm 1}, where $\bld{x}_{r}(t)$, $\dot{\bld{x}}_{r}(t)$, and $\ddot{\bld{x}}_{r}(t)$ represent the position, velocity, and acceleration of a reference trajectory. The payload, assumed as a mass point in the robot's TCP reference frame, is denoted by $m_{TCP}$. To outline the algorithm, line 4 updates the payload in $robot(\cdot)$, containing all kinematic, inertial, and topological robot data. Line 5 initiates a time loop with final value $M$, where lines 7-9 compute the first and second-order differential inverse kinematics based on the reference trajectory.
The manipulator Jacobian ($\bld{J}$) \cite{lynch2017modern} and its first time derivative ($\dot{\bld{J}_{\;}}\!\!$) are computed in lines 14-25. By following a screw theory notation \cite{Bib:selig,lynch2017modern}: $\bld{s}_i$ is the screw vector of the $i$-th actuator expressed at local coordinates, $\mbox{Ad}_{G_0^{i}}$ is the adjoint operator parameterized by the homogeneous transformation matrix $\bld{G}_0^{i}\in SE(3)$ that represents the motion of frame $i$ with respect to $0$ frame, $\bld{J}^{s}$ is the spatial Jacobian, $\dot{\bld{J}^{s}}$ is its time derivative and the subscript $[:,i]$ refers to their $i$-th column. Also $\bld{r}_{0}^{TCP}$ is the distance vector from $0$ reference frame to $TCP$ frame and $\mbox{ad}_{s_i}$ stands for the adjoint operator that performs a Lie bracket, see \cite{paz2024analytical} for a detailed form of these adjoints operators. The superscript $+$ stands for the generalized pseudoinverse. Line 10 computes the robot's inverse dynamics \cite{Bib:Featherstone} based on $\bld{q}$, its first two time derivatives $\dot{\bld{q}}$ and $\ddot{\bld{q}}$, and the $robot(\cdot)$ structure \cite{zhu2010virtual}\cite{petrovic2022mathematical}. The sensitivity metrics for this work are represented by $\bld{\Psi}$ which is computed in line 11 and it is a function of the linear velocities and forces of the actuators denoted by $\bld{v}_{\!x}$ and $\bld{f}_{\!x}$, respectively.
\subsection{Simulation and Results}
\label{sec:simulation}
The sensitivity analysis is conducted on a 3-DoF HDMM to monitor the behaviour of the performance metrics with respect to the payload. The parallel-serial manipulator under study is illustrated in Fig. \ref{fig:Mobile_platform}, which is composed by two closed-kinematic chains and one telescope mechanism. The manipulator is forced to track a spiral reference trajectory that covers most of its operational space in a time of $M=8\pi$ (s). We vary the payload $m_{TCP}$ from 0 to 200 (kg) and perform the PD of $\bld{\Psi}$ with respect to $m_{TCP}$ by means of finite differences with a payload perturbation factor of $\Delta_m = 2\!\times\!10^{-24}$ (kg), as \eqref{eq:PD_formula}.
\begin{equation}
    \frac{\partial \bld{\Psi}}{\partial m_{TCP}} \ \approx \ \frac{\bld{\Psi}(m_{TCP} + \Delta_m ) - \bld{\Psi}(m_{TCP})}{\Delta_m}
    \label{eq:PD_formula}
\end{equation}
The four scalar sensitivity metrics under study are defined as follows, computed and stacked in the variable $\bld{\Psi}$ for each EMLA at lift, tilt, and telescope, as itemized below:
\begin{itemize}
    \item \textit{Delivered power}: It is the power retrieved from the EMLA at an instant of time at the load side of actuator:
    \begin{equation}
        \psi_{1i} = v_{x_i} f_{x_i}
        \label{equation:minimumeffort}
    \end{equation}
    where $v_{x_i}$ and $f_{x_i}$ are the scalar linear velocity and force of the $i$-th actuator. The evaluation of the power metric ($\boldsymbol{\psi_1}$) and its PD ($\partial\boldsymbol{\psi_1} / \partial m_{TCP}$) are depicted in Fig. \ref{fig:Power}.
    \begin{figure}[H]
        \centering
        \centerline{{\includegraphics[width=9cm]{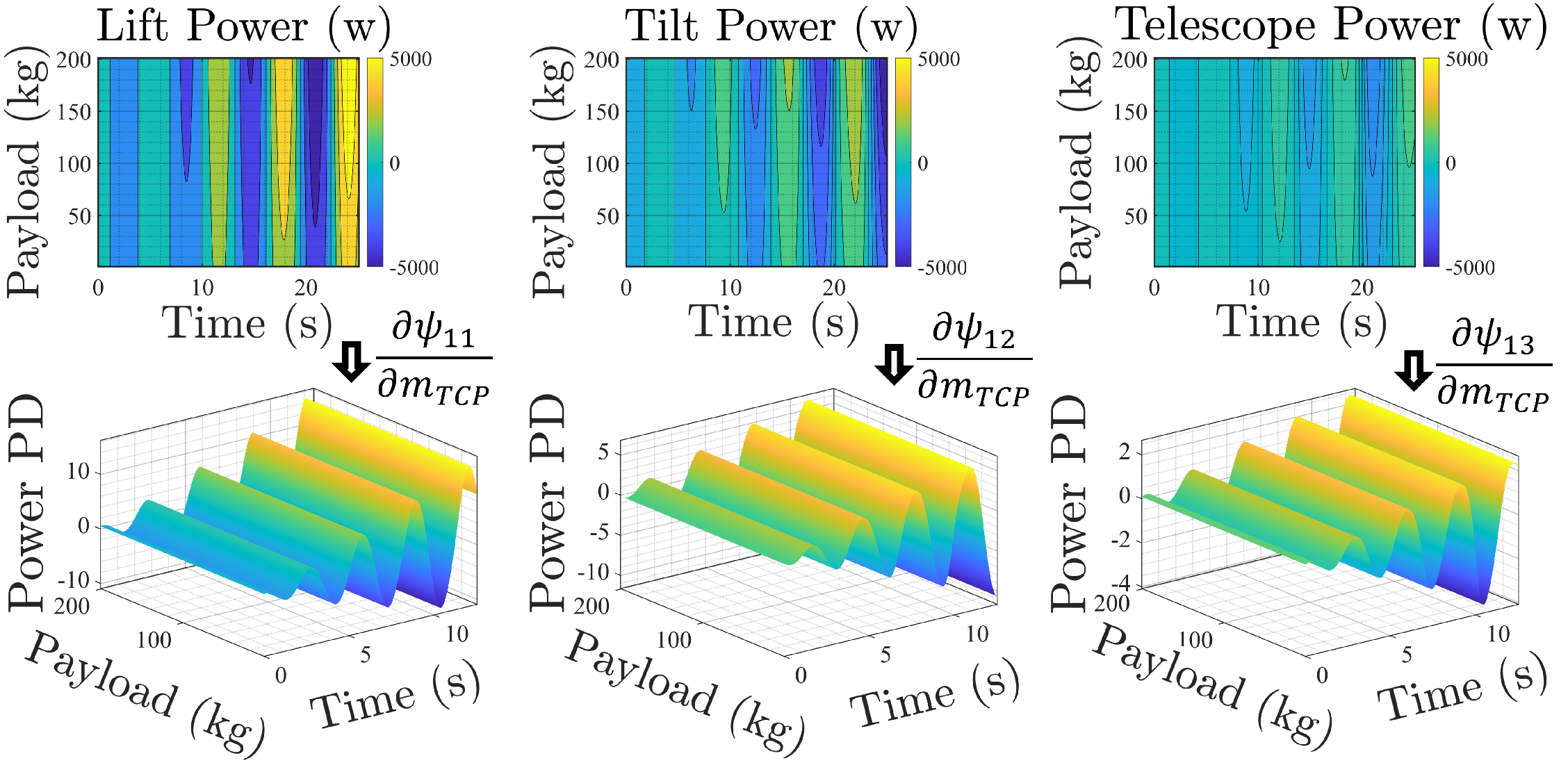}}}
        \caption{Delivered power and its PD with respect to TCP payload in 3-DoF HDMM joints over time} 
        \label{fig:Power}
    \end{figure}
    \item \textit{Actuators' linear force}: The linear force exerted by the EMLA in the direction of motion at the load side:   
    \begin{equation}
        \psi_{2i} = f_{x_i}
        \label{equation:minimumforce}
    \end{equation}
    The assessment of the force metric ($\boldsymbol{\psi_2}$) and its PD ($\partial\boldsymbol{\psi_2} / \partial m_{TCP}$) are shown in Fig. \ref{fig:Force}.
    \begin{figure}[H]
        \centering
        \centerline{{\includegraphics[width=9cm]{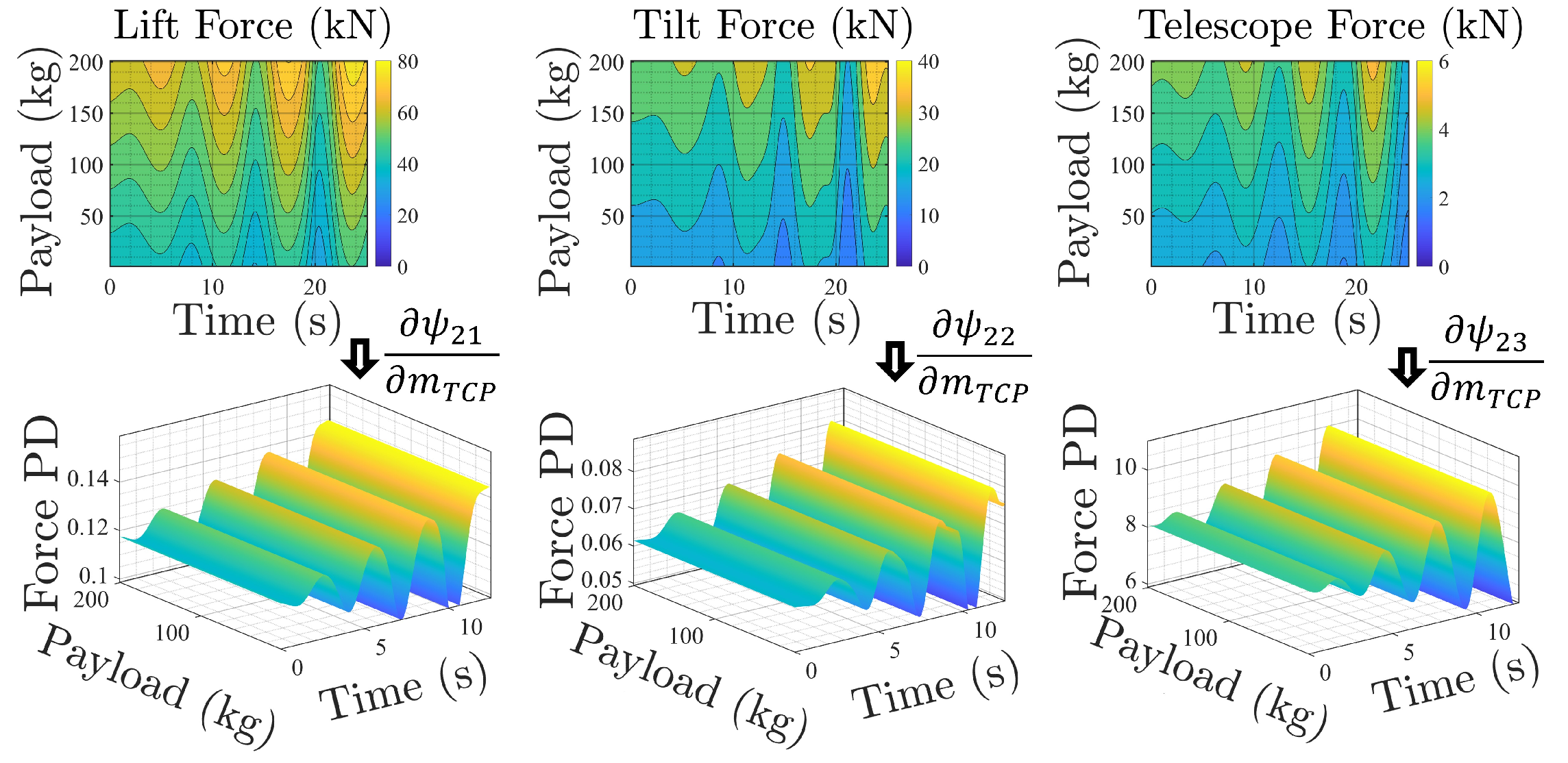}}}
        \caption{Linear Force and its PD with respect to TCP payload in 3-DoF HDMM joints over time} 
        \label{fig:Force}
    \end{figure}
    \item \textit{Energy expenditure}: The total energy consumed by the EMLA over a period, calculated as the sum of the output power divided by the efficiency, integrated over time:
    \begin{equation}
        \psi_{3i} = \Delta_t \sum_{t=0}^{M} \frac{v_{x_i} f_{x_i}}{\eta_{\text{EMLA}_i} (f_{x_i} , v_{x_i})}
        \label{equation:energyexpenditure}
    \end{equation}
    where $\Delta_t$ is the increment of time. Fig. \ref{fig:Energy} illustrates the energy metric ($\boldsymbol{\psi_3}$) and its PD with respect to the TCP payload ($\partial\boldsymbol{\psi_3} / \partial m_{TCP}$).
    \begin{figure}[H]
        \centering
        \centerline{{\includegraphics[width=7.3cm]{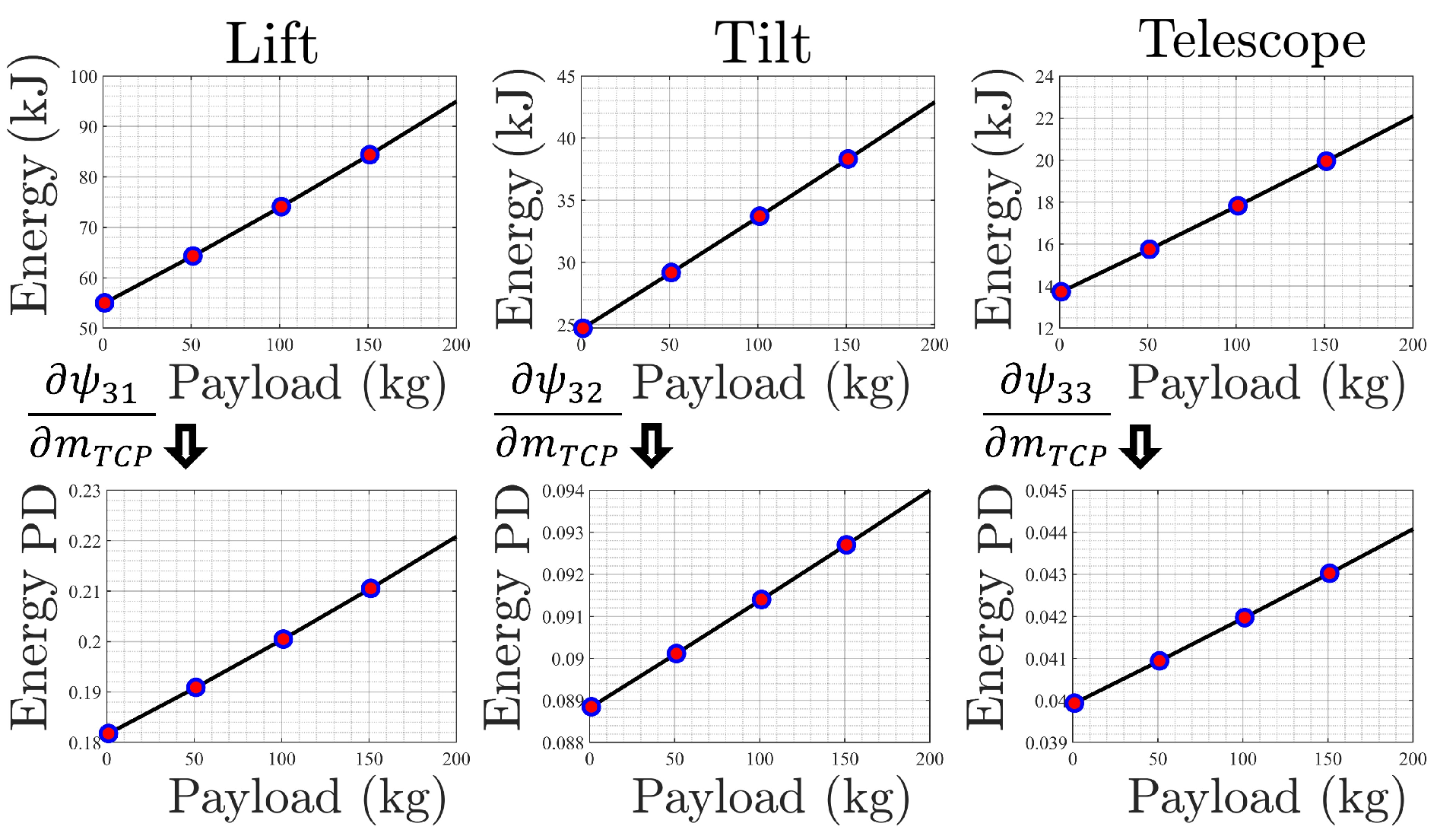}}}
        \caption{Energy consumption and its PD with respect to TCP payload in 3 HDMM joints} 
        \label{fig:Energy}
    \end{figure}
    \item \textit{Efficiency}: The efficiency of the EMLAs, calculated as the ratio of the output mechanical power to the input electrical power at an instant of time:   
    \begin{equation}
        \psi_{4i} = \frac{v_{x_i} f_{x_i}}{\sqrt{3} V_{\!L\!L} I_{\!L\!L} \cos(\phi)}
        \label{equation:efficiency}
    \end{equation}
    Fig. \ref{fig:Efficiency} presents the efficiency metric ($\boldsymbol{\psi_4}$) and its PD with respect to the TCP payload ($\partial\boldsymbol{\psi_4} / \partial m_{TCP}$).
    \begin{figure}[tbp]
        \centering
        \centerline{{\includegraphics[width=7.3cm]{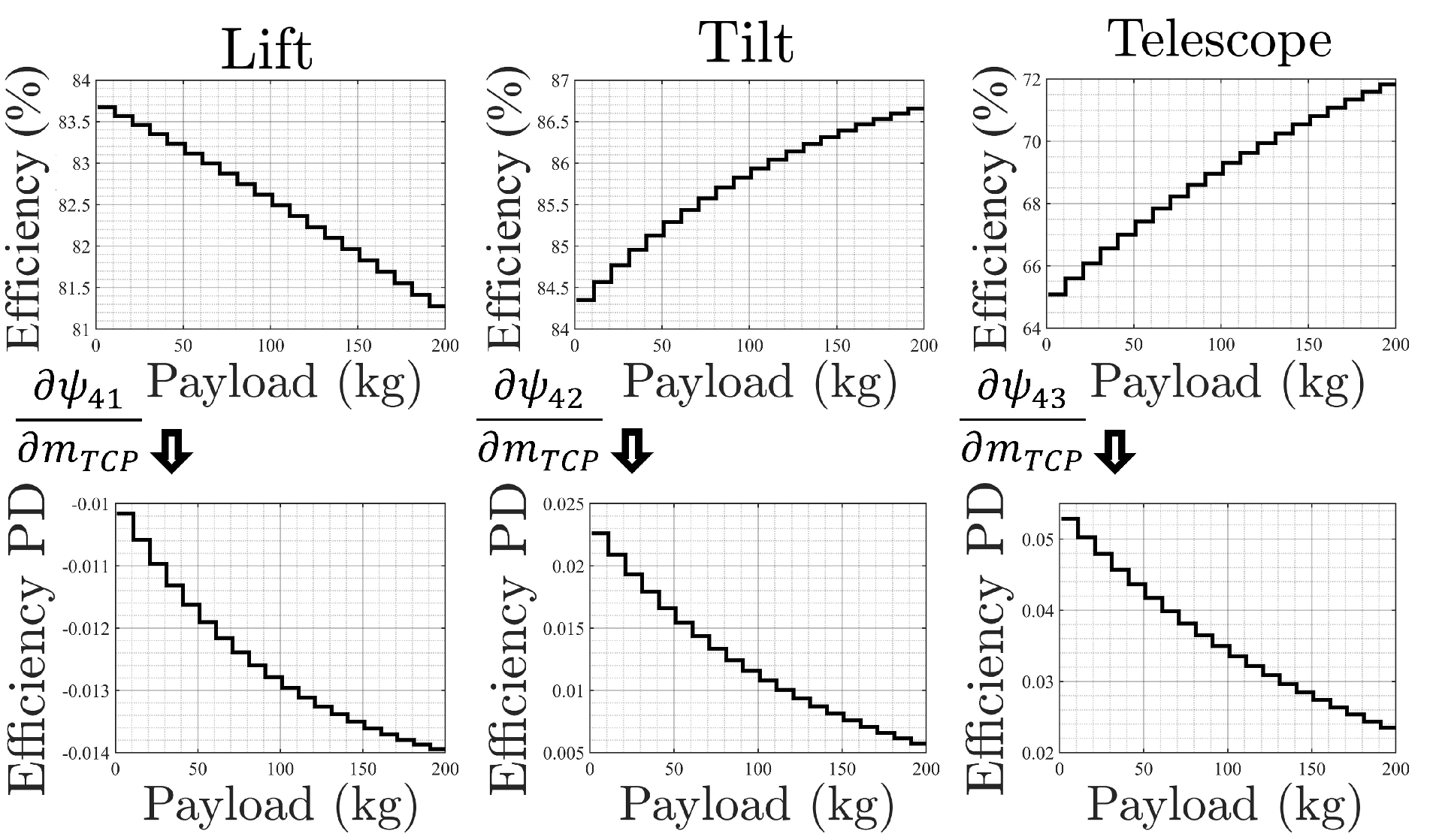}}}
        \caption{Efficiency of EMLAs and its PD with respect to TCP payload in 3 HDMM joints} 
        \label{fig:Efficiency}
    \end{figure}
    
\end{itemize}
\begin{table}[H]
  \centering
  \renewcommand{\arraystretch}{0.9} 
  \begin{tabular}{p{0.9\linewidth}} 
    \hline
    \\
    \multicolumn{1}{c}{\textbf{Algorithm 1}: Second-Order Inverse Differential Kinematics} \\
    \\
    \hspace{0.4cm}\textbf{Input}: $\bld{x}_{r}(t)$, $\dot{\bld{x}}_{r}(t)$, $\ddot{\bld{x}}_{r}(t)$, $m_{TCP}$.\\
    \hspace{0.4cm}\textbf{Output}: $\bld{\Psi}$.\\
    \\
    {\small 1}\hspace{0.2cm}$\bld{\Psi} \leftarrow$ \textbf{getMetrics}($\bld{x}_{r}(t)$, $\dot{\bld{x}}_{r}(t)$, $\ddot{\bld{x}}_{r}(t)$, $m_{TCP}$)\\
    {\small 2}\hspace{0.2cm}\textbf{getMetrics}($\bld{x}_{r}(t)$, $\dot{\bld{x}}_{r}(t)$, $\ddot{\bld{x}}_{r}(t)$, $m_{TCP}$)\\
    {\small 3}\hspace{0.6cm}Initialize $\bld{q}$\\
    {\small 4}\hspace{0.6cm}Update payload $robot(m_{TCP})$\\
    {\small 5}\hspace{0.6cm}\textbf{For} $t=0:M$ \textbf{then}\\
    {\small 6}\hspace{1.0cm}$[\bld{J}(t)$, $\dot{\bld{J}_{\;}}\!\!(t)] \leftarrow$ \textbf{getJacobian}($\bld{q}(t)$, $\dot{\bld{q}}(t)$)\\
    {\small 7}\hspace{1.0cm}$\dot{\bld{q}}(t)=\bld{J}^{+}(\bld{q}(t))\dot{\bld{x}}_{r}(t)$\\
    {\small 8}\hspace{1.0cm}Update $\bld{q}(t)$\\
    {\small 9}\hspace{1.0cm}$\ddot{\bld{q}}(t)=\bld{J}^{+}(\bld{q}(t))[\ddot{\bld{x}}_{r}(t) - \dot{\bld{J}_{\;}}\!\!(\bld{q}(t),\dot{\bld{q}}(t))]\dot{\bld{q}}(t)$\\
    {\small 10}\hspace{0.85cm}$[\bld{v}_{\!x}(t)$, $\bld{f}_{\!x}(t)] \leftarrow$ \textbf{RNEA}($\bld{q}(t)$, $\dot{\bld{q}}(t)$, $\ddot{\bld{q}}(t)$, $robot(m_{TCP})$)\\
    {\small 11}\hspace{0.85cm}$\bld{\Psi} \leftarrow \textbf{evaluateMetrics} (\bld{v}_{\!x}(t), \bld{f}_{\!x}(t))$\\
    {\small 12}\hspace{0.5cm}\textbf{end}\\
    {\small 13}\hspace{0.1cm}\textbf{end}\\
    {\small 14}\hspace{0.1cm}\textbf{getJacobian}($\bld{q}(t)$, $\dot{\bld{q}}(t)$)\\
    {\small 15}\hspace{0.5cm}$\dot{\mbox{Ad}}_{G_0^{0}} \leftarrow \bld{0}$\\
    {\small 16}\hspace{0.5cm}\textbf{For} $i=1:n$ \textbf{then}\\
    {\small 17}\hspace{0.85cm}$\bld{J}_{\![:,i]}^{s} = \mbox{Ad}_{G_0^{i}}\bld{s}_i$\\
    {\small 18}\hspace{0.85cm}$\dot{\mbox{Ad}}_{G_0^{i}} = \dot{\mbox{Ad}}_{G_0^{\lambda}}\mbox{Ad}_{G_{\lambda}^{i}} + \mbox{Ad}_{G_0^{i}}\mbox{ad}_{s_i}\dot{q}_i$\\
    {\small 19}\hspace{0.85cm}$\dot{\bld{J}^{s}}_{\![:,i]} = \dot{\mbox{Ad}}_{G_0^{i}}\bld{s}_i$\\
    {\small 20}\hspace{0.5cm}\textbf{end}\\
    {\small 21}\hspace{0.5cm}$\bld{r}_{0}^{TCP} \ \leftarrow \ \bld{G}_{0}^{n}\bld{G}_{n}^{TCP}$\\
    {\small 22}\hspace{0.5cm}$\bld{J} = \begin{bmatrix}
		\bld{I} & -[\bld{r}_{0}^{TCP}]
		\end{bmatrix}\bld{J}^{s}$\\
    {\small 23}\hspace{0.5cm}$\dot{\bld{r}}_{0}^{TCP} \ = \ \bld{J} \dot{\bld{q}}$\\
    {\small 24}\hspace{0.5cm}$\dot{\bld{J}_{\;}} = \begin{bmatrix}
		\bld{0} & -[\dot{\bld{r}}_{0}^{TCP}]
		\end{bmatrix}\bld{J}^{s} + \begin{bmatrix}
		\bld{I} & -[\bld{r}_{0}^{TCP}]
		\end{bmatrix}\dot{\bld{J}^{s}}$\\
    {\small 25}\hspace{0.1cm}\textbf{end}\\
    \\
    \hline
  \end{tabular}
\end{table}
 %
%
\section{Conclusions}
This study has presented a sensitivity analysis of a PMSM-powered EMLA-driven HDMM, focusing on how varying payload masses impact performance metrics such as power, force, energy expenditure, and efficiency. The findings offer crucial insights into the behavior of these actuators under different loading conditions, providing an insight for the development of next-generation, fully-electrified HDMMs.

The key contributions of this research are:
\begin{itemize}
    \item Introducing a mathematical model of a gear-equipped PMSM-powered EMLA. This model encompasses equivalent circuit representations for the inverter, EM, gearbox, screw mechanism, and load. By employing the analogy of force-to-voltage and speed-to-current, this framework facilitates a thorough understanding of the governing dynamics and enables precise performance analysis of the EMLA.
    \item Assisting in the informed selection of EMLAs for each joint of the manipulator by analyzing the sensitivity of required force and delivered power with respect to the TCP payloads.
    \item Facilitating the optimal sizing of batteries required to ensure effective operation of electric HDMM, thereby enhancing their performance and efficiency.
\end{itemize}

These contributions are vital for appropriate selection of actuators and battery capacity, and paves the way for the design of high-performance, autonomous HDMMs capable of efficient and sustainable operation. Future research could expand this sensitivity analysis to encompass a broader range of manipulator designs and operational conditions.
\vspace{12pt}
	\bibliography{biblio}
	\bibliographystyle{unsrt}
\end{document}